\documentstyle[12pt]{article}
\nopagebreak

\def\eq{\begin{equation}}

\def\ee{\end{equation}}

\def\eqa{\begin{eqnarray}}

\def\eea{\end{eqnarray}}

\begin{document}


\centerline{\Large{\bf Photon electroproduction 
off nuclei}}

\medskip

\centerline{\Large{\bf in the $\Delta$-resonance region}}

\vskip 1.5cm

\centerline{\large{B.~Pasquini}}

\vskip 1.0cm

\centerline{\small Dipartimento di Fisica Nucleare e Teorica, Universit\`a
di Pavia, and}

\centerline{\small Istituto Nazionale di Fisica Nucleare, 
Sezione di Pavia, Pavia, Italy}

\vskip 1.5cm


\begin{abstract}

\noindent 
The cross section for the $A(e,e'\gamma)A$ reaction is calculated, 
investigating the contribution from the nuclear target with respect to the 
radiative corrections from the electron. The reaction mechanism is studied for
photon emission in the $\Delta$-resonance region, varying the scattering 
geometry and analyzing the most favourable kinematical conditions to extract
information on the nuclear system.

\end{abstract}

\bigskip



$\Delta$-resonance 

\vskip 1.5cm
\clearpage

\begin{section}{Introduction}
The photon electroproduction process off nuclei is potentially a very useful 
tool to investigate the nuclear structure. In the one photon exchange 
approximation, the reaction is described by the coherent sum of the 
Bethe-Heitler (BH) amplitude and the full virtual Compton scattering (FVCS)
amplitude. The first process describes the emission of bremsstrahlung
photons by the electron in the nuclear electromagnetic field and is exactly 
calculable from QED. The second process corresponds to electron 
scattering by exchange of a virtual photon which is scattered by the nucleus
into a real final photon and is given by a linear combination of virtual 
Compton scattering (VCS) amplitudes.
The competition between the two processes makes a very difficult task to extract
 experimentally the interesting information on the nuclear response involved 
in the reaction.
Such experiments have been proposed in the past in the energy range of the 
giant resonances~\cite{[Hubbard],[Acker]}, exploring the best kinematical 
conditions to extract information on the first excited nuclear levels.
VCS in  the same energy region has been also considered 
in ref.~\cite{[ad]} within a new formulation in terms of nuclear 
generalized polarizabilities and an experiment has been performed to study the 
$4,44$ MeV excited state $(J^{\pi}=2^+)$ in $^{12}$C~\cite{[Papanicolas]}.

Recently, new interest has emerged to study the nucleon structure through 
VCS~\cite{[Farrar],[Guichon]}.
Experiments have already been scheduled at MAMI-B~\cite{[Auditm]} below
the pion production threshold in order to explore the non perturbative 
structure of the nucleon, while proposed experiments at CEBAF~\cite{[Auditc]}
will extend the measurements to higher virtual photon momenta.

In this paper, we propose to investigate the same reaction mechanism in 
the case of a nuclear target, discussing the possibility to extract relevant 
information on the nuclear structure at intermediate energies.
VCS in the $\Delta$-resonance region has already been discussed 
in ref.~\cite{[Pasquini2]}, focusing the attention on the new 
accessible information with respect to the real Compton scattering. 
The discussion is now extended to include the evaluation of photon
bremsstrahlung contribution of the electrons,
exploring the interplay between the BH and the FVCS amplitude
in different scattering geometries and trying to disentangle as clearly as 
possible the pure nuclear contribution.    

In sect. 2, the explicit expression for the cross section of the 
$A(e,e'\gamma)A$ reaction is given, separating the contribution 
from the BH, the FVCS and the interference terms. Results for $^4$He 
at the electron energies accessible at MAMI and CEBAF are
presented in sect. 3 and concluding remarks are reported in the final section.
\end{section}

\begin{section}{Cross section of photon electroproduction 
off nuclei}

To describe the photon electroproduction reaction off nuclei
\eq
\begin{array}{ccccccccc}
e & + & A &\rightarrow & e' & + & A & + & \gamma , \\
  &   &   &            &    &   &   &   &\\
(h,k) &   & (p) &            & (h',k') &   & (p')&   
&(\varepsilon^{\mu}_{\lambda'},q'), 
\end{array}
\ee
we choose to work in the laboratory frame system and denote
with $h,$ $h'$ and $k^{\mu}=(E,\vec k)$, $k'^{\mu}=(E',\vec{ k}')$ 
the helicity and the 
four-momentum of the initial and final electron, respectively.
Neglecting the nucleus recoil, the four-momentum of the 
nucleus in the initial and final state is $p^{\mu}=p'^{\mu}=(M_A,0)$, while 
$\varepsilon^{\nu}_{\lambda'}(q'^{\mu})$ and 
$q'^{\mu}=(\omega '=q',\vec q')$ are 
the polarization vector and four-momentum of the final 
photon, respectively.

Without observing polarization effects, the differential cross section 
is given by
\eq
{\rm d}\sigma\equiv
\frac{{\rm d}^5\sigma}{{\rm d}\Omega_{e'}{\rm d}\Omega_{\gamma}{\rm d}\omega'}=
\frac{m_e^2}{(2\pi)^5}\frac{\omega'}{2} 
\frac{\vert\vec k'\vert}{\vert\vec k\vert}
\sum_{h', \lambda '}\overline{\sum_{h}}\vert M^{{\rm BH}}_{h'\lambda',h}+
M^{{\rm FVCS}}_{h'\lambda',h}\vert^2,
\ee
where $m_e$ is the electron mass and $M^{{\rm BH}}_{h'\lambda',h}$ and
$M^{{\rm FVCS}}_{h'\lambda',h}$ are the helicity amplitudes for the BH and 
 FVCS processes, respectively.
Due to the coherence of the two effects, the cross section is the sum of 
three terms 
\eq {\rm d}\sigma= {\rm d}\sigma^{{\rm BH}}+ {\rm d}\sigma^{{\rm FVCS}}+
{\rm d}\sigma^{{\rm INT}},
\ee
where ${\rm d}\sigma^{{\rm BH}}$ and ${\rm d}\sigma^{{\rm FVCS}}$ are
the contributions of photon emission by the electron and the nucleus, 
while ${\rm d}\sigma^{{\rm INT}}$ is obtained from 
the interference between the BH and FVCS amplitudes. 

The BH amplitude is evaluated in the external field approximation, where the 
nucleus is treated as source of a static external Coulomb field, which 
transfers to the electron the momentum $\vec{\kappa}=\vec p '+\vec q'
-\vec p.$  
The explicit expression for the BH amplitude reads
\eq
M^{{\rm BH}}_{h'\lambda',h}=-i\,e^2A^{\mu}(\vec{\kappa})
j^{{\rm BH}}_{\mu\nu}(h,h')\varepsilon^{\nu}_{\lambda'}(q'^{\mu}),
\ee
where $A^{\mu}(\vec{\kappa})=(F(\vec{\kappa})/\vert\vec{\kappa}\vert^2,0,0,0)$
 is the momentum-space Coulomb potential corresponding to the nuclear charge 
form factor $F(\vec{\kappa})$ and $j^{BH}_{\mu\nu}(h,h')$ is 
the leptonic current for the process~\cite{[Rohrlich]}.
Finally, from the squared modulus of the BH amplitude the well known
Bethe-Heitler formula for ${\rm d}\sigma^{BH}$~\cite{[Rohrlich]}
is derived.

In the one photon exchange approximation, the FVCS amplitude is given by a 
linear combination of VCS amplitudes
\eq
M^{{\rm FVCS}}_{h'\lambda',h}=\frac{ie}{Q^2}\sum_{\lambda}(-1)^{\lambda}
\varepsilon^{\mu \,*}_{\lambda}(q)j_{\mu}(h,h')
M^{{\rm VCS}}_{\lambda'\lambda},
\ee
where $j_{\mu}(h,h')$ is the electron current,
$\varepsilon^{\mu}_{\lambda}(q^{\mu})$ 
and 
$q^{\mu}=k^{\mu}-k'^{\mu}=(\omega,\vec q)$ are the polarization vector and 
four-momentum of the virtual intermediate photon, respectively and 
$Q^2=q^2-\omega^2.$

$M^{{\rm VCS}}_{\lambda'\lambda}$ is the model-dependent nuclear transition 
amplitude, describing the scattering off nucleus of a virtual photon 
with helicity $\lambda=0\pm 1$ 
into a real photon with helicity $\lambda'=\pm 1$. 
At intermediate energies, this term
is dominated by the resonant contribution describing the excitation of the 
$\Delta(1232)$ isobar inside the nucleus.
As explained in details in ref.~\cite{[Pasquini2],[Pasquini]}, the 
$\Delta$-resonance 
contribution can be satisfactorily evaluated within a local-density approximation
to the $\Delta$-hole model. 
A background non-resonant contribution can be included, taking into account the
seagull diagram and the scattering amplitude due to s-wave 
pion production and absorption on a single nucleon.
Within the framework of this model, the FVCS cross section is directly 
calculated from the expression
\eq
{\rm d}\sigma^{{\rm FVCS}}=\frac{e^2}{(2\pi)^5}\frac{\omega '}{8 Q^4}
\frac{\vert\vec k'\vert}{\vert\vec k\vert}\,
\sum_{\lambda, \lambda',\bar{\lambda}}
(M_{\lambda'\lambda}^{{\rm VCS}})\,
L_{\lambda\bar{\lambda}}\,
(M_{\lambda'\bar{\lambda}}^{{\rm VCS}})^{*},
\label{eq:crosssec} 
\ee
where
$L_{\lambda\bar{\lambda}}$ is the lepton tensor (see, e.g.,
ref.~\cite{[PhysRep]}). 

The summation in eq.~(\ref{eq:crosssec}) can be explicitly written in terms of
four structure functions of the nucleus
\eqa
& &\sum_{\lambda, \lambda',\bar{\lambda}}
(M_{\lambda'\lambda}^{{\rm VCS}})
L_{\lambda\bar{\lambda}}\,
(M_{\lambda'\bar{\lambda}}^{{\rm VCS}})^{*}
=\qquad\qquad\qquad\nonumber\\
& &
=L_{00} W_{\rm L} + L_{11}\, 2 W_{\rm T} + 
L_{01} W_{\rm LT} \cos\alpha  + L_{1-1}\, 2 W_{\rm TT} \cos 2\alpha, 
\eea
where $\alpha$ is the azimuthal angle of the emitted photon with respect to
the electron scattering plane.
The pure transverse structure function $W_{\rm T}$ is the incoherent sum of 
photon-helicity flip and non-flip contributions, while the transverse-transverse
 $W_{{\rm TT}}$ gives the interference contribution from helicity flip and 
non-flip amplitudes.
The pure longitudinal response is given by the structure function $W_{{\rm L}}$ 
and longitudinal-transverse interference contributions are contained in 
$W_{{\rm LT}}.$
With respect to the real Compton scattering, new information are available
through the $W_L$ and $W_{LT}$ responses, while the same $W_T$ and $W_{TT}$
structure functions can now be explored varying indipendently the energy and 
momentum of the incoming photon.

The VCS amplitudes come into play also to determine the interference 
contribution ${\rm d}\sigma^{{\rm INT}}.$ Separating the leptonic and nuclear 
terms, one has
\eq
{\rm d}\sigma^{{\rm INT}}=
\frac{m_e^2\omega'}{(2\pi)^5}\,
\frac{\vert\vec k'\vert}{\vert\vec k\vert}\,
\frac{F(\vec{\kappa})e^3}{\vert\vec{\kappa}\vert ^2 Q^2}\,
{\rm Re}\Big\{\sum_{\lambda,\lambda'}L^{{\rm INT}}_{\lambda\lambda'}
(M^{\rm VCS}_{\lambda'\lambda})^*\Big\},
\ee
where now the helicity amplitudes $M^{{\rm VCS}}_{\lambda'\lambda}$
are linearly combined by the tensor
$L^{{\rm INT}}_{\lambda\lambda'}$,
derived by the interference between the leptonic currents in the BH and FVCS 
amplitude.
\end{section}

\begin{section}{Results}

The total information on the nuclear dynamics
is summarized 
in the nuclear structure functions of eq.~(\ref{eq:crosssec}).  
To access experimentally these responses, we face the problem to find
those kinematical conditions where the bremsstrahlung contributions of the 
electron are as small as possible.
The relevant variables  to determine the scattering geometry are given
by
\eq
(E,E',\theta_{e'},q,\omega,\theta_{\gamma},\alpha),
\ee
where $\theta_{e'}$ is the scattering angle of the outgoing electron with
respect to the initial electron and $\theta_{\gamma}$ is the scattering angle 
of the final photon with respect to the virtual one.
Neglecting the recoil of the nucleus, the nuclear structure functions depend 
only on the variables $\omega=\omega',$ $\theta_{\gamma}$ and $q.$ 
Since we are interested in the nuclear dynamics in the $\Delta$-resonance 
region, we keep fixed $\omega=310$ MeV and explore the nuclear responses as a 
function of $\theta_{\gamma}$ in different regions of $q.$
For given values of $q$ and $\omega$, the electron variables $E,$$E'$ and
$\theta_{e'}$ are related by two conditions
\eq
E-E'=\omega,
\ee
\eq
\vert\vec{q}^{\,2}\vert=E^2+E'^2-2EE'\cos\theta_{e'}.
\ee
As a consequence, we have at our disposal only one indipendent variable, which 
we choose to be the incoming electron energy E.   

The azimuthal angle $\alpha$ of the outgoing photon is choosen in a such way to 
separate the different contributions of the structure functions in the total 
cross section. Measurements of the cross section at complementary angles 
$\alpha$ can be combined to define the two quantities
\eqa
A_+(\alpha)&=&{\rm d}\sigma(\alpha)+{\rm d}\sigma(180^{{\rm o}}-\alpha),
\nonumber\\
& &\\
A_-(\alpha)&=&{\rm d}\sigma(\alpha)-{\rm d}\sigma(180^{{\rm o}}-\alpha).
\nonumber
\eea
From eq.~(\ref{eq:crosssec}), we deduce that in the left-right asymmetry 
$A_-$  the $W_{{\rm LT}}$ contribution to the FVCS cross section is singled out.
Since out-of-plane kinematics is in general favoured to reduce the BH 
contaminations~\cite{[Auditm],[Auditc]}, we will take the two values
$\alpha=45^{{\rm o}}$ and $\alpha=135^{{\rm o}}.$ 
On the other hand, these conditions allow to estimate the relative contribution
of the $W_{{\rm L}}$ and $W_{{\rm T}}$ structure functions in the $A_+$
combination. The importance of the pure longitudinal response is completely
negligible with respect to $W_{{\rm T}}$~\cite{[Pasquini2]} and 
in the regions of small BH contaminations we could extract from the $A_+$ 
measurement direct information on the $W_{{\rm T}}$ structure function.

A different analysis has been recently performed in ref.~\cite{[Gomez]},
where the interplay between the BH and FVCS contributions is investigated
in the $\Delta$-resonance region at different values of $E$ and 
$\theta_{e'}$ and 
integrating the cross section over the photon azimuthal angle.
With the choice of $(E, \, q,\, \alpha)$ as the set of indipendent variables, 
we can now examine more closely the role of the nuclear responses, emphasizing
the new information available with respect to real Compton scattering from
the behaviour as a function of the momentum transfer $q.$

In figs. 1-3, results for $A_+(45^{{\rm o}})$ are shown as a function of the 
photon scattering angle $\theta_{\gamma}$, separating the BH and 
FVCS contributions with dotted and dot-dashed lines, respectively. The 
calculations are performed for $^4$He at three different values of the incoming 
electron energy ($E=500,\,885$ and $2000$ MeV), keeping fixed 
$\omega=310$ MeV and investigating the region of the virtual photon momentum 
$q$ between $330$ and $480$ MeV.
BH contaminations are dominant at forward angles of the  outgoing photon, 
becoming negligible in the backward region. In particular, keeping 
the electron energy fixed at $E=500$ MeV,  the 
nuclear contribution can be better seen at small values of $q,$ 
where it becomes the leading term for $\theta_{\gamma}\geq 60^{{\rm o}}.$ 
On the other hand, increasing the electron 
energy, we observe the same trend as a function of $q$, while the angular 
region where FVCS dominates is extended to smaller $\theta_{\gamma}$ 
(for instance, $\theta_{\gamma}\geq40^{{\rm o}}$ at $E=2000$ MeV 
and $q=330$ MeV).

Due to the angular dependence on $(1-\cos \theta_{\gamma})$ and
$(1+\cos \theta_{\gamma})$ of the helicity flip and non-flip amplitudes, 
respectively, the $W_{{\rm T}}$ structure function is dominated at large angle
by the helicity flip contribution. This term is not yet well understood
on the theoretical point of view~\cite{[Pasquini],[KMO]} and important
information in a large range of $q$ could be clearly extracted  from VCS
experiments.

Results for the left-right asymmetry $A_-(\alpha=45^{{\rm o}})$ are plotted in 
figs. 4-6 for the same kinematics previously explained. 
The $W_{{\rm LT}}$ contributions are given in absolute value and
the minima correspond to a sign change from positive to negative values.
As expected, the small longitudinal term is very hard to extract, even if 
the overwhelming BH contribution falls down in the backward scattering region.

The most interesting situation to investigate is at the electron energy 
$E=2000$ MeV.
In the $q$ range between $ 330$ and $430$ MeV and at large scattering angles,
the BH term becomes negligible or is completely canceled by the negative 
$W_{{\rm LT}}$ contribution and the left-right asymmetry is 
controlled by the positive interference between the FVCS and the BH amplitudes.
On the other hand, in the interference cross section 
${\rm d}\sigma^{{\rm INT}}$ the VCS amplitudes are linearly summed, with 
a negligible importance of the longitudinal part with respect to the transverse
 one. At the highest $q=480$ MeV, the BH contribution is the only important
term and becomes dominant even at smaller 
values of $q$ for decreasing electron energy. 

More complex combinations of the total cross section can be analyzed to extract 
also the $W_{{\rm TT}}$ contribution (for instance, by subtracting twice the 
cross section at $\alpha=90^{{\rm o}}$ from  $A_+(45^{{\rm o}})$ ), but the 
 small values of the involved cross sections should require too 
high-precision experiments.
\end{section}

\begin{section}{Concluding remarks}
 The cross section for the $A(e,e'\gamma)A$ reaction has been calculated in 
the $\Delta$-resonance region, investigating the interplay between the BH and 
FVCS amplitudes for various scattering geometries.
Out-of-plane kinematics, with $\alpha=45^{{\rm o}}$ and $\alpha=135^{{\rm o}}$,
 has been choosen to disentangle the role of the $W_{{\rm LT}}$ and 
$W_{{\rm T}}$ structure functions in the plus and minus combinations of
the total cross sections. 

In order to explore the nuclear responses as a 
function of the momentum $q$ in the range between $330$ and $480$ MeV, the
most favourable case occurs for high energy of the incoming electron. 
The pure transverse $W_{{\rm T}}$ can be clearly separated from BH 
contaminations at backward scattering angles, where new interesting information
could be obtained on the helicity flip amplitude contributions.

More difficult appears to explore the longitudinal nuclear response.
On one hand, in the $W_{{\rm LT}}$ structure function the small longitudinal
contribution is amplified by the interference with the transverse term, but we
have to compete with the overwhelming BH contaminations in the measurement of 
the left-right asymmetry.  
On the other hand, in the regions where the BH and FVCS interference is
pronounced, the role of the longitudinal amplitude in 
${\rm d}\sigma^{{\rm INT}}$ is obscured by the transverse terms.

\end{section}


\clearpage


\centerline{\bf Figure captions}

\medskip

Fig. 1. $A_+$ combination of the differential cross section for photon
electroproduction off $^4$He, calculated at $\alpha=45^{{\rm o}}$ as a function
of the photon scattering angle $\theta_{\gamma}$ for the incoming electron 
energy $E=500$ MeV.
The transfer energy $E-E'$ is fixed at $310$ MeV and the virtual photon momentum
is taken at the different values $q=330,$ $380,$ $430$ and $480$ Mev.
The dashed and dot-dashed lines are the separate contributions of the BH
and FVCS processes, respectively, while the solid line is obtained from the 
coherent sum of the two contributions.    

\smallskip

Fig. 2. The same as in fig. 1 but for $E=885$ MeV.

\smallskip

Fig. 3. The same as in fig. 1 but for $E=2000$ MeV.

\smallskip

Fig. 4. $A_-$ asymmetry calculated in the same kinematical conditions as in 
fig. 1.

\smallskip

Fig. 5. The same as in fig. 4 but for $E=885$ MeV.

\smallskip

Fig. 6. The same as in fig. 4 but for $E=2000$ MeV.

\smallskip

\end{document}